\documentclass[aps,11pt,final,
notitlepage, oneside,  nobibnotes, nofootinbib,
superscriptaddress, showpacs, centertags, showkeys, assume]{revtex4}

\usepackage{epsfig}
\usepackage{graphicx}
\usepackage{amsfonts}
\usepackage{amsmath}
\begin{document}

\title{Rigorous pion electromagnetic form factor behavior in the spacelike region}

\author{M.Beli\v cka}
\address{Dept. of Theoretical Physics, Comenius University, Bratislava,
Slovak Republic}
\author{S.Dubni\v cka}
\address{Institute of Physics, Slovak Academy of Sciences,
Bratislava, Slovak Republic}
\author{A.Z.Dubni\v ckov\'a}
\address{Dept. of Theoretical Physics, Comenius University, Bratislava,
Slovak Republic}
\author{A. Liptaj}
\address{Institute of Physics, Slovak Academy of Sciences,
Bratislava, Slovak Republic}

 \vspace{2cm}
\begin{abstract}
   New precise experimental information on $\sigma_{tot}(e^+e^- \to
\pi^+ \pi^-)$ is transferred into the space-like region, by taking
advantage of the analyticity. As a result a rigorous pion
electromagnetic form factor behavior in the spacelike region is obtained. The latter with
some existing model predictions is compared.
\end{abstract}
\pacs{12.40.Vv, 13.40.Hq, 13.66.Be, 13.88.+e} \keywords{QCD, pion
form factor, analyticity}

\maketitle

The pion electromagnetic (EM) form factor (FF) $F_{\pi}(Q^2)$ with
the squared four-momentum transfer $Q^2=-t$ is one of the most
simple objects of investigations in strong interaction physics. In
spite of this fact, there is no theory till now able to explain
all its known features. Even in the space-like region, where the
pion EM FF behavior is expected to be represented by a simple
smooth decreasing curve between the norm $F_{\pi}(0)=1$ and the
pQCD asymptotic behavior \cite{1}-\cite{3}
\begin{equation}
F_{\pi}(Q^2)_{Q^2\to \infty}\sim
\frac{64\pi^2f_{\pi}^2}{(11-2/3n_f)Q^2\ln Q^2/\Lambda^2},
\label{eq1}
\end{equation}
all known attempts (see \cite{4}-\cite{8}) to reach experimentally
measurable region give not uniform results.

   In this paper, it is shown how pion EM FF can be reconstructed in
the space-like region  with the help of the accurate data on the
total cross-section $\sigma_{tot}(e^+e^- \to \pi^+ \pi^-)\equiv
\sigma_{tot}(t)$ in the elastic $4m_{\pi}^2\leq t \leq
(m_{\pi^0}+m_{\omega})^2$ region, which plays a dominant role in
our prediction. On the basis of Phragmen-Lindel\"of theorem, the
assumption is made that asymptotically pion EM FF in the Minkowski
region has an analogous form to that in the Euclidean one. As a
result, the asymptotic form of the imaginary part of pion EM FF in
the time-like region is found to be helpful to specify correct
parameterization of corrections from the interval $(m_{\pi^0}+
m_{\omega})^2\leq t \leq +\infty $. All these ingredients are
linked up together via dispersion integrals and as a result, a
prediction for the pion EM FF in the space-like region can be
achieved.

   Really, the analytic properties of the pion EM FF, by means of the
Cauchy formula and assuming the validity of the pQCD asymptotic
behavior (\ref{eq1}) in all directions of the complex $t$-plane,
can be transformed into the dispersion relation without any
subtractions
\begin{equation}
F_{\pi}(Q^{2})=\frac{1}{\pi}\int\limits_{4m_{\pi}^{2}}^{t_{\pi^0\omega}}
  \frac{Im^E\,F_{\pi}(t')}{t'+Q^{2}}dt'
  +\frac{1}{\pi}\int\limits_{t_{\pi^0\omega}}^{\infty}
  \frac{Im^A\,F_{\pi}(t')}{t'+Q^{2}}dt'.\label{eq2}
\end{equation}

   Using the normalization condition $F_\pi(0)=1$ for $Q^2=0$ in (\ref{eq2}) one gets the sum
rule for the pion FF imaginary parts
\begin{equation}
1=\frac{1}{\pi}\int\limits_{4m_{\pi}^{2}}^{t_{\pi^0\omega}}
  \frac{Im^E\,F_{\pi}(t')}{t'}dt'
  +\frac{1}{\pi}\int\limits_{t_{\pi^0\omega}}^{\infty}
  \frac{Im^A\,F_{\pi}(t')}{t'}dt'.\label{eq3}
\end{equation}

   Another super-convergence sum rule for the same imaginary
parts, namely
\begin{equation}
0=\frac{1}{\pi}\int\limits_{4m_{\pi}^{2}}^{t_{\pi^0\omega}}
  {Im^E\,F_{\pi}(t')}dt'
  +\frac{1}{\pi}\int\limits_{t_{\pi^0\omega}}^{\infty}
  Im^A\,F_{\pi}(t')dt',\label{eq4}
\end{equation}
can be derived by an application of the Cauchy theorem to
$F_\pi(t)$ in the complex t-plane and its pQCD asymptotics
(\ref{eq1}).

   In all previous three integral relations, we have automatically separated
the  elastic region $4m_\pi^2\leq t \leq (m_{\pi^0} + m_\omega)^2$
contributions of $Im^E\,F_{\pi}(t)$ (therefore superscript E)
which for $F_\pi(Q^2)$ at $Q^2=-t=0$ represents up to 90\% of the
total $Im\,F_{\pi}(t)$= $Im^E\,F_{\pi}(t)$ + $Im^A\,F_{\pi}(t)$,
as one can see from further considerations.

   In order to evaluate the first integral in (\ref{eq2}), one can
apply the following method of extracting $Im^E\,F_{\pi}(t)$ from
$\sigma_{tot}(e^+e^-\to \pi^+\pi^-)\equiv \sigma_{tot}(t)$, which
is the foremost quantity for obtaining  of experimental values of
the pure isovector pion EM FF in the time-like region.

   As the electron-positron annihilation into two charged pions is
of the EM nature, one can treat it in the one-photon-exchange
approximation and as a result, there are no model ingredients in
the extraction of $|F_{\pi}(t)|$ from the measured cross-section.
Since two final-state pions with total orbital moment $l=1$ (due
to the spin of the photon) have the isospin $I=1$ and a positive
$G$-parity, the pion EM FF is of the pure isovector nature and all
resonances to be seen in the pion EM FF data can be only
isovectors (the $\rho$-meson family) with $G=+1$ and with all
other quantum numbers of the photon, like $J=1$ and negative
intrinsic and charge parities.

   Nevertheless, in Review of Particle Physics \cite{9} one finds
also isoscalar vector meson isospin violating decays into two
charged pions, $\omega(782) \to \pi^+\pi^-$ with fraction
$(\Gamma_i/\Gamma)=1.53\%$ and $\Phi(1020) \to \pi^+\pi^-$ with
fraction $(\Gamma_i/\Gamma)=7.3\times 10^{-5}\%$, which contribute
through higher order corrections to the $e^+e^- \to \pi^+\pi^-$
process and experimentalists are unable to eliminate them from
final results.

   In order to obtain the pure isovector pion EM FF experimental
information from existing data on $e^+e^- \to \pi^+\pi^-$ process,
we write its total cross-section in the form

\begin{equation}
  \sigma_{tot}(t)=\frac{\pi\alpha^{2}\beta_\pi^{3}}{3t}\mid F_{\pi\rho}(t)+
  \xi\cdot\exp(i\alpha)F_{\pi\omega}(t)\mid^2; \quad
  \beta_{\pi}=[(t-4m_{\pi}^2)/t]^{1/2},\label{eq5}
\end{equation}
where $F_{\pi\rho}(t)$ and $F_{\pi\omega}(t)$ represent $\rho$-
and $\omega$- meson contributions to the $e^+e^- \to \pi^+\pi^-$
process, respectively, and the $\Phi$-meson contribution is
neglected, as we are interested only in $\sigma_{tot}(e^+e^- \to
\pi^+\pi^-)$ in the elastic region $4m_\pi^2\leq t \leq
(m_{\pi^0}+m_{\omega})^2$. The so-called $\rho$-$\omega$
interference amplitude $\xi$ can be expressed through the partial
decay width $\Gamma(\omega \to \pi^+\pi^-)$ by the relation
\begin{equation}
  \xi=\frac{6}{\alpha m_{\omega}}\big (\frac
  {m_{\omega}^2}{m_{\omega}^2-4m_{\pi}^2}\big )^{3/4}[\Gamma({\omega}\to
  e^+e^-)\cdot\Gamma({\omega}\to\pi^+\pi^-)]^{1/2},\label{eq6}
\end{equation}
and the $\rho$-$\omega$ interference phase $\alpha$ is
\begin{equation}
\alpha=\arctan{\frac{m_\rho\Gamma_\rho}{m_\rho^2 -m_\omega^2}}.
\label{eq7}
\end{equation}

   Because the $\omega$-vector meson is a very narrow resonance, one
can approximate the $\omega$-meson contribution to $e^+e^- \to
\pi^+\pi^-$ process in (\ref{eq5}) by the Breit-Wigner form

\begin{equation}
   F_{\pi\omega}(t)=\frac{m_{\omega}^2}{m_{\omega}^2-t-im_{\omega}\Gamma_{\omega}}\label{eq8}.
\end{equation}

Further, first we exploit the pion FF phase representation
$F_{\pi\rho}(t)=|F_\pi(t)|\cdot exp(i\delta_{\pi})$ in (\ref{eq5})
and subsequently the pion FF phase identity
$\delta_{\pi}\equiv\delta^1_1$ with the P-wave isovector $\pi\pi$
scattering phase shift $\delta^1_1(t)$ for $4m_\pi^2\leq t \leq
(m_{\pi^0} + m_{\omega})^2$. The latter follows just from the
elastic pion FF unitarity condition $Im^E\,F_{\pi}(t)$=$\mid
F_{\pi}(t)\mid$$
e^{i\delta_{\pi}(t)}$$e^{-i\delta^1_1(t)}$$\sin{\delta^1_1(t)}$.
 As a result the quadratic equation for the absolute
value of pure isovector pion FF data is obtained \cite{10}
\begin{equation}
\mid F_{\pi\rho}(t)\mid^2+2Z(t)\mid F_{\pi\rho}(t)\mid
  +\left\{\frac{\xi^2m_{\omega}^4}{(m_{\omega}^2-t)^2+m_{\omega}^2\Gamma_{\omega}^2}
  -\frac{3t}{\pi\alpha^2\beta_{\pi}^3}\sigma_{tot}(t)\right\}=0\ ,\label{eq9}
\end{equation}
with the physical solution
\begin{equation}
\mid F_{\pi\rho}(t)\mid= -Z(t)+ \left
\{Z^2(t)+\frac{3t}{\pi\alpha^2\beta_{\pi}^3}\sigma_{tot}(t)-
  \frac{\xi^2m_{\omega}^4}{(m_{\omega}^2-t)^2+m_{\omega}^2\Gamma_{\omega}^2}\right \}^{1/2}\label{eq10}
\end{equation}
and
\begin{equation}
  Z(t)=\frac{\xi m_{\omega}^2}{(m_{\omega}^2-t)^2+m_{\omega}^2\Gamma_{\omega}^2}
  [(m_{\omega}^2-t)\cos(\alpha-\delta_1^1)
  -m_{\omega}\Gamma_{\omega}\sin(\alpha-\delta_1^1)].\label{eq11}
\end{equation}
The data on  $Im^E F_{\pi\rho}(t)$ with errors for $4m_\pi^2\leq t
\leq (m_{\pi^0} + m_{\omega})^2$ are then determined by the
relation
\begin{equation}
 Im^E\,F_{\pi\rho}(t)= \mid F_{\pi\rho}(t)\mid \sin{\delta^1_1}, \label{eq12}
\end{equation}
using experimental information on $\xi$, $\alpha$, $m_\omega$,
$\Gamma_\omega$, the recently measured up data in Frascati
\cite{11} by the radiative return and in Novosibirsk \cite{12,13}
improved experimental information on $\sigma_{tot}(e^+e^-\to
\pi^+\pi^-)$ and the suitable parameterization \cite{10} of
$\delta^1_1(t)$. Then the first integral of (\ref{eq2}) as a
function of $Q^2$ is a smoothly decreasing curve and the first
integrals of (\ref{eq3}) and (\ref{eq4}) give
\begin{equation}
\frac{1}{\pi}\int_{4m_\pi^2}^{t_{\pi^0\omega}}\frac{Im^E\,F_{\pi}(t')}{t'}dt'=
0.8995\label{eq13}
\end{equation}
and
\begin{equation}
\frac{1}{\pi}\int_{4m_\pi^2}^{t_{\pi^0\omega}}Im^E\,F_{\pi}(t')dt'=
0.5023,\label{eq14}
\end{equation}
respectively, where we have already identified $Im^E
F_{\pi\rho}(t)$ with $Im^E F_{\pi}(t)$.

   In order to estimate the second integral in (\ref{eq2}) as a
function of $Q^2$, one has to know something about the asymptotic
$Im^A\,F_\pi(t)$ for $(m_{\pi^0} + m_{\omega})^2\leq t < +\infty$.
The analytic continuation of (\ref{eq1}) to the upper boundary of
the pion FF cut on the positive real axis of  the $t=-Q^2$ plane
leads to the pion FF imaginary part
\begin{equation}
Im\,F_{\pi}(t)_{t \to \infty}\sim
-\pi\frac{(64\pi^2f_{\pi}^2)}{(11-2/3n_f)t\ln^2t/\Lambda^2}.
\label{eq15}
\end{equation}
The positivity of all data on $Im^E\,F_\pi(t)$ following from
(\ref{eq12}) for $4m_\pi^2\leq t\leq (m_{\pi^0} + m_{\omega})^2$
and the asymptotic form (\ref{eq15}) can be satisfied
simultaneously only if $Im^A\,F_\pi(t)$ in (\ref{eq2})-(\ref{eq4})
acquires at least one zero value at $t_z$ for $t>(m_{\pi^0} +
m_{\omega})^2$ and vanishes asymptotically from the negative
values as $t \to +\infty$. The simplest function reflecting all
these required properties is
\begin{equation}
 Im^A\,F_\pi(t)= \pi\frac{64\pi^2f_{\pi}^2}{(11-2/3n_f)}\frac{t_z-t}{(t-C)^2\ln^2{t/\Lambda^2}},\label{eq16}
\end{equation}
with the parameter values
\begin{equation}
\Lambda=0.7226~GeV, \quad C=-9.7255~GeV^2, \quad t_z=4.6975~
GeV^2,\quad n_f=13.2517 \label{eq17}
\end{equation}
to be determined from conditions
\begin{eqnarray}
  Im^E\, F_\pi(t)_{|_{t=t_{\pi^0\omega}}}&=&  Im^A\, F_\pi(t)_{|_{t=t_{\pi^0\omega}}}\nonumber \\
 \frac{d}{dt}Im^E\, F_\pi(t)_{|_{t=t_{\pi^0\omega}}}&=& \frac{d}{dt} Im^A\, F_\pi(t)_{|_{t=t_{\pi^0\omega}}}\label{eq18} \\
 0.1005&=&\frac{64\pi^2f_\pi^2}{(11-2/3n_f)}\int\limits_{t_{\pi^0\omega}}^{\infty}\frac{t_z-t'}{t'(t'-C)^2\ln^2{t'/
 \Lambda^2}}dt' \nonumber \\
 -0.5023&=&\frac{64\pi^2f_\pi^2}{(11-2/3n_f)}\int\limits_{t_{\pi^0\omega}}^{\infty}\frac{t_z-t'}{(t'-C)^2\ln^2{t'/
 \Lambda^2}}dt'\nonumber
\end{eqnarray}
obtained by using also the values of the integrals (\ref{eq13})
and (\ref{eq14}), respectively.

   The pion EM FF space-like region behavior calculated by the
dispersion relation (\ref{eq2}) is displayed in Fig.1 (solid
line), where also recent theoretical predictions \cite{4}-\cite{7}
are presented for comparison.

\begin{figure}[t,h,b]
    \centering
    \scalebox{0.6}{\includegraphics{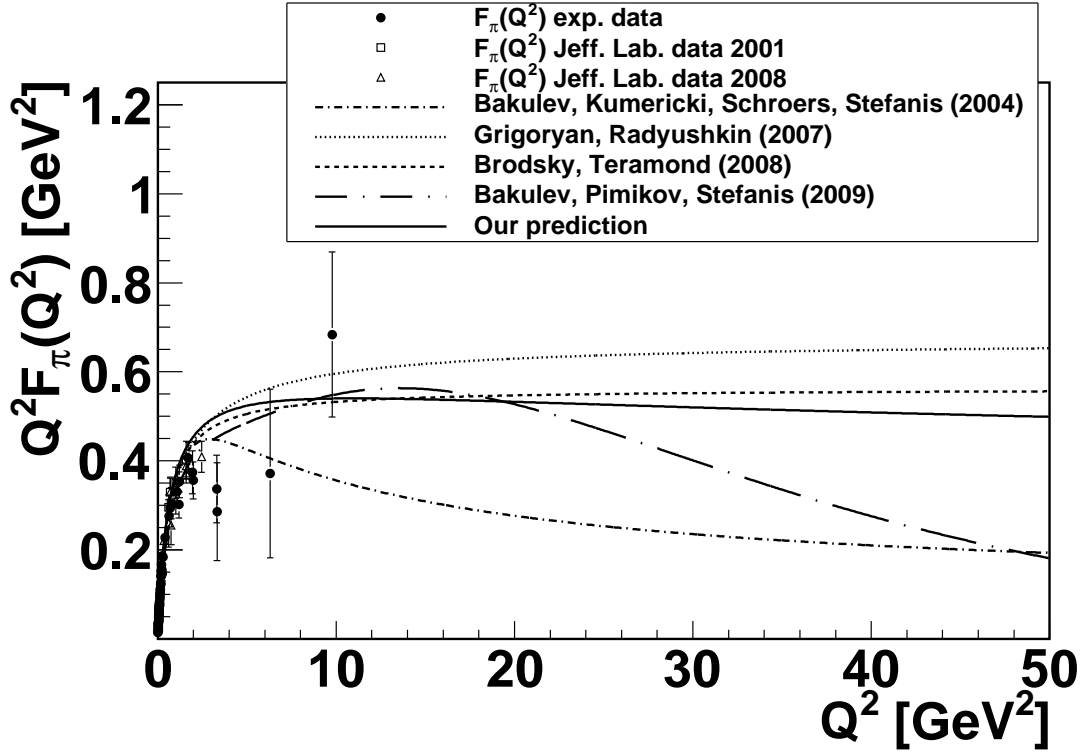}}
    \caption{\small{Theoretical predictions of the pion EM FF behavior in the space-like region and their
                     comparison with existing data.}}
    \label{fig:1}
\end{figure}

\begin{figure}[t,h,b]
    \centering
    \scalebox{0.6}{\includegraphics{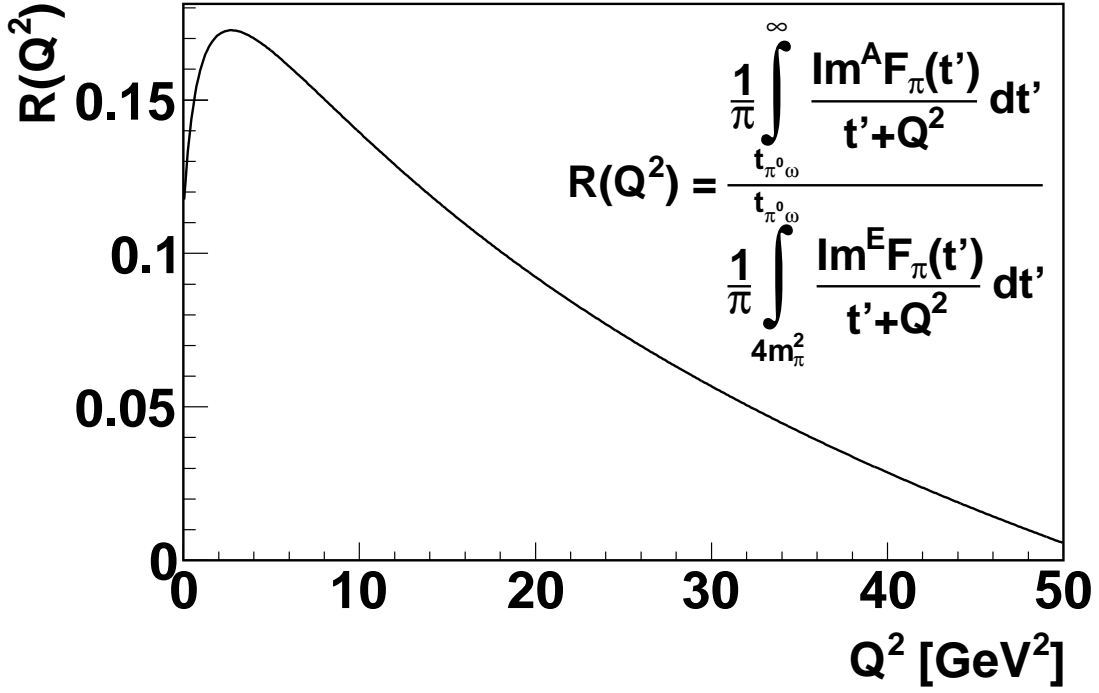}}
    \caption{\small{The ratio of the $Im^AF_\pi(t)$ contribution to the $Im^EF_\pi(t)$ one in predicted
                    pion EM FF behavior in the space-like region.}}
    \label{fig:2}
\end{figure}

   In Fig.2 we draw the ratio of the second integral in
(\ref{eq2}) to the first one as a function of $Q^2$ in order to
demonstrate our approach to be more or less model independent.
Really, as one can clearly see from Fig.2, the correction of the
weakly model dependent parametrization (\ref{eq16}) of the
$Im^AF_\pi(t)$ for $(m_{\pi^0}+m_\omega)^2\leq t < +\infty$
becomes negligible with increased values of $Q^2$. As a result,
our prediction of the pion EM FF in Fig.1 with increased values of
$Q^2$ is more and more model independent.

   We defend the reliability of our prediction for the pion EM FF in the space-like
region, presented in Fig.1,  also by a prediction of the complex
pion EM FF on the upper boundary of the cut in the time-like
region and compare it with existing data. In this region,
predictions seem to be more sensitive to the analytic
approximations and the issues discussed above and their comparison
with the accurate time-like data surely will be more severe
self-consistent test of the whole elaborate approach.

   We start with the dispersion relation

\begin{equation}
F_{\pi}(t)=\frac{1}{\pi}\lim_{\varepsilon \to 0}
\{\int\limits_{4m_{\pi}^{2}}^{t_{\pi^0\omega}}
  \frac{Im^E\,F_{\pi}(t'+i\varepsilon)}{t'-t-i\varepsilon}dt'
  +\int\limits_{t_{\pi^0\omega}}^{\infty}
  \frac{Im^A\,F_{\pi}(t'+i\varepsilon)}{t'-t-i\varepsilon}dt'\},\label{eq19}
\end{equation}
where the first integral in brackets is singular if the interval
$4m_\pi^2\leq t\leq (m_{\pi^0} + m_{\omega})^2$ is considered and
the second integral in brackets is singular if the complex pion EM
FF is calculated within the interval $t_{\pi^0\omega}<t<+\infty$ .

   For an evaluation of the singular integrals, one can use the well known
symbolic so-called Sokhotsky-Plemelj formula from the theory of
functions of complex variables

\begin{equation}
\lim_{\varepsilon \to 0} \frac{1}{t'-t\mp
i\varepsilon}=\textsl{P}\frac{1}{t'-t}\pm i\pi \delta(t'-t).
\label{eq20}
\end{equation}

Then considering the first integral in (\ref{eq19}) to be singular
one practically obtains
\begin{equation}
\frac{1}{\pi}\lim_{\varepsilon \to 0}
\int\limits_{4m_{\pi}^{2}}^{t_{\pi^0\omega}}
\frac{Im^E\,F_{\pi}(t'+i\varepsilon)}{t'-t-i\varepsilon}dt'=\frac{1}{\pi}\textsl{P}
\int\limits_{4m_{\pi}^{2}}^{t_{\pi^0\omega}}
\frac{Im^E\,F_{\pi}(t')}{t'-t}dt'+i\int\limits_{4m_{\pi}^{2}}^{t_{\pi^0\omega}}
Im^E\,F_{\pi}(t')\delta(t'-t)dt',\label{eq21}
\end{equation}
where $\textsl{P}$ denotes that the Cauchy principal value
\begin{equation}
\frac{1}{\pi}\textsl{P}
\int\limits_{4m_{\pi}^{2}}^{t_{\pi^0\omega}}
\frac{Im^E\,F_{\pi}(t')}{t'-t}dt'=\frac{1}{\pi}\lim_{\delta \to 0}
\{\int\limits_{4m_{\pi}^{2}}^{t-\delta}
\frac{Im^E\,F_{\pi}(t')}{t'-t}dt'
+\int\limits_{t+\delta}^{t_{\pi^0\omega}}
\frac{Im^E\,F_{\pi}(t')}{t'-t}dt'\}\equiv Re^E\,
F_{\pi}(t)\label{eq22}
\end{equation}
has to be taken and the second integral in (\ref{eq21}) gives just
$Im^E\,F_{\pi}(t)$, by means of which the dominant part of the
pion EM FF space-like behavior is found.

\begin{figure}[t,h,b]
    \centering
    \scalebox{0.6}{\includegraphics{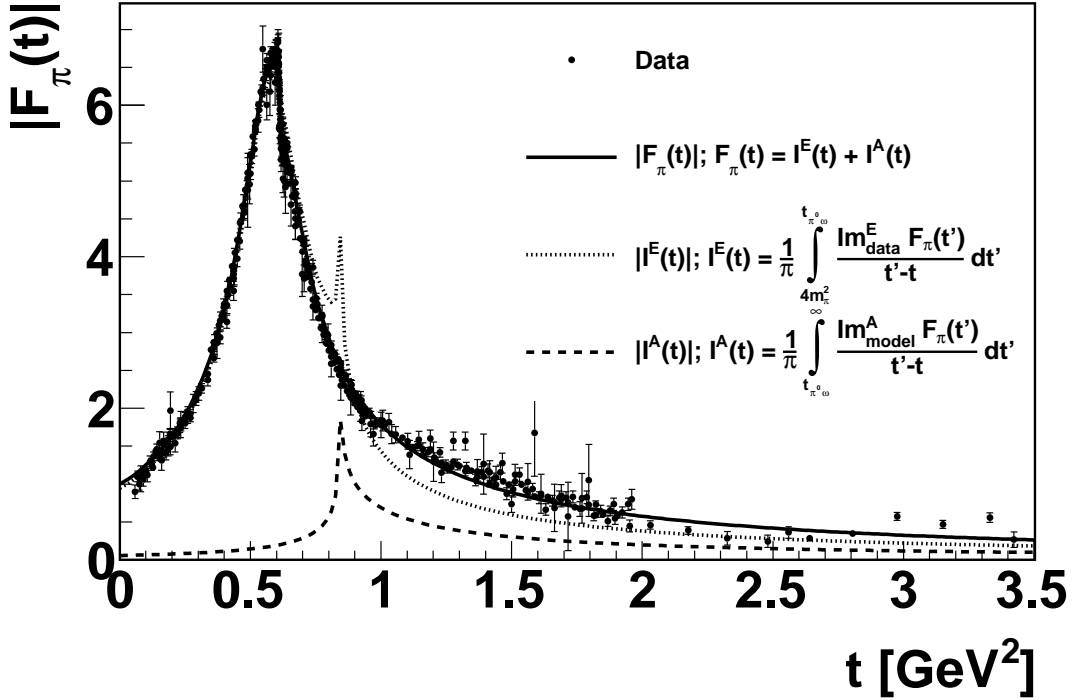}}
    \caption{\small{A self-consistent reconstruction of the absolute value of
                    the pion EM FF behavior in the time-like region
                    with the help of the accurate experimental
                    information on $\sigma_{tot}(e^+e^- \to
                    \pi^+\pi^-).$}}
    \label{fig:3}
\end{figure}

   In a similar way a contribution of the second singular integral
in (\ref{eq19}) to the complex pion EM FF on the upper boundary of
the cut in the time-like region can be evaluated.

   Numerical predictions for the absolute values of both integrals
in (\ref{eq19}), as well as the absolute value of the whole
complex pion EM FF in the time-like region and its comparison with
existing data up to $t=3.5 GeV^2$ is presented in Fig.3.

   A righteous agreement (see Fig.3) of the predicted absolute value of
the pion EM FF in the time-like region with existing data confirms
a reliability of our prediction of $F_{\pi}(t)$ in the space-like
region as it is presented in Fig.1.

   The authors would like to thank E.Betak for reading of the manuscript.
The support of the Slovak Grant Agency for Sciences VEGA under
Grant No.2/0009/10 is acknowledged.

\end{document}